\documentclass [11pt,a4paper] {article}

\usepackage[cp1252]{inputenc}
\usepackage{amssymb}
\usepackage{amsmath}
\usepackage{amsfonts,amssymb}
\usepackage[dvips]{graphicx}
\usepackage{bbm}
\usepackage{enumerate}
\usepackage{amsthm}
\usepackage{cancel}

\DeclareMathAlphabet{\mathpzc}{OT1}{pzc}{m}{it}

\def\SmallColSep{\setlength{\arraycolsep}{1pt}}

\begin{document}

\title{Algebraic structures identified with bivalent and non-bivalent semantics of experimental quantum propositions}

\author{Arkady Bolotin\footnote{$Email: arkadyv@bgu.ac.il$\vspace{5pt}} \\ \textit{Ben-Gurion University of the Negev, Beersheba (Israel)}}

\maketitle

\begin{abstract}\noindent The failure of distributivity in quantum logic is motivated by the principle of quantum superposition. However, this principle can be encoded differently, i.e., in different logico-algebraic objects. As a result, the logic of experimental quantum propositions might have various semantics. E.g., it might have either a total semantics, or a partial semantics (in which the valuation relation – i.e., a mapping from the set of atomic propositions to the set of two objects, 1 and 0 – is not total), or a many-valued semantics (in which the gap between 1 and 0 is completed with truth degrees). Consequently, closed linear subspaces of the Hilbert space representing experimental quantum propositions may be organized differently. For instance, they could be organized in the structure of a Hilbert lattice (or its generalizations) identified with the bivalent semantics of quantum logic or in a structure identified with a non-bivalent semantics. On the other hand, one can only verify – at the same time – propositions represented by the closed linear subspaces corresponding to mutually commuting projection operators. This implies that to decide which semantics is proper – bivalent or non-bivalent – is not possible experimentally. Nevertheless, the latter allows simplification of certain no-go theorems in the foundation of quantum mechanics. In the present paper, the Kochen-Specker theorem asserting the impossibility to interpret, within the orthodox quantum formalism, projection operators as definite $\{0,1\}$-valued (pre-existent) properties, is taken as an example. The paper demonstrates that within the algebraic structure identified with supervaluationism (the form of a partial, non-bivalent semantics), the statement of this theorem gets deduced trivially.\\

\noindent \textbf{Keywords:} Truth value assignment; Hilbert lattice; Invariant-subspace lattices; Quantum logic; Supervaluationism; Many-valued semantics; Kochen-Specker theorem.\\
\end{abstract}

\section{Introduction}  

\noindent To understand quantum mechanics from \emph{a logico-algebraic perspective}, an assignment of truth values to \emph{experimental propositions} – i.e., meaningful declarative sentences that are (or make) statements about a physical system – plays an essential role. Let us elucidate this point.\\

\noindent Assume that any experimental proposition associated with a quantum system is represented by a closed linear subspace of the Hilbert space $\mathcal{H}$ characterizing the system. If all results of unperformed experiments on the quantum system are classically pre-determined, then each not-yet-proven experimental proposition pertaining to the system is either true or false.\\

\noindent Suppose that $A$ and $B$ are \emph{compatible} experimental propositions, i.e., ones that can be proven together. This implies that they allow an attribution of truth values to their logical conjunction, $A \sqcap B$. In line with \cite{Fine}, let us introduce \emph{the product and sum rules} for such propositions, namely,\smallskip

\begin{equation} \label{PROD} 
   {[\mkern-3.3mu[
      A
      \sqcap
      B
   ]\mkern-3.3mu]}_v
   =
   {[\mkern-3.3mu[
      A
   ]\mkern-3.3mu]}_v
   \times
   {[\mkern-3.3mu[
      B
   ]\mkern-3.3mu]}_v
   \;\;\;\;  ,
\end{equation}
\\[-40pt]

\begin{equation} \label{SUM} 
   {[\mkern-3.3mu[
      A
      \sqcup
      B
   ]\mkern-3.3mu]}_v
   =
   {[\mkern-3.3mu[
      A
   ]\mkern-3.3mu]}_v
   +
   {[\mkern-3.3mu[
      B
   ]\mkern-3.3mu]}_v
   -
   {[\mkern-3.3mu[
      A
      \sqcap
      B
   ]\mkern-3.3mu]}_v
   \;\;\;\;  ,
\end{equation}
\smallskip

\noindent where ${A}\sqcup{B}$ stands for the logical disjunction of $A$ and $B$, the double-bracket notation denotes \emph{a valuation} \cite{Dalen, Dunn}, that is, a mapping from the set of atomic propositions, symbolized by $\mathbb{P}$, to the Boolean domain $\mathbb{B}_{2}$ (the set of truth values, \emph{true} and \emph{false}, renamed to 1 and 0, correspondingly), i.e.,\smallskip

\begin{equation} \label{VAL} 
   v
   \mkern-3.3mu
   :
   \mkern2mu
   \mathbb{P}
   \to
   \mathbb{B}_{2}
   \;\;\;\;  ,
\end{equation}
\smallskip

\noindent such that for each proposition of $\mathbb{P}$, say $P$, one has $v(P) = {[\mkern-3.3mu[P]\mkern-3.3mu]}_v$.\\

\noindent Now, recall that in accordance with the Kochen-Specker theorem \cite{Kochen}, it is impossible to assign truth values to all experimental propositions associated with a quantum system in a consistent manner. In detail, this theorem (usually called \emph{the KS theorem}) asserts that there always exists a set $S$ of compatible experimental propositions about the quantum system characterized by $N$-dimensional Hilbert space $\mathcal{H}$, such that all propositions in $S$ cannot have truth values satisfying the rules (\ref{PROD}) and (\ref{SUM}).\\

\noindent The KS theorem was proved on the set $L(\mathcal{H})$ of the closed linear subspaces of $\mathcal{H}$ such that each pair of the subspaces in $L(\mathcal{H})$ has a meet (greatest lower bound) corresponding to their set-theoretic intersection. Since the original demonstration for $|S|=117$ and $N=3$, more and more proofs of the KS theorem have been found for the same or higher $N$ but lesser $|S|$ (see \cite{Peres, Kernaghan, Cabello, Yu}, to cite but a few examples), and the task has become to reduce the technical difficultness required to prove the KS theorem in order to make the issues involved clearer.\\

\noindent Then again, from the logico-algebraic perspective, the set of the closed linear subspaces of $\mathcal{H}$, in which all pairs have a meet, forms an algebra called \emph{the Hilbert lattice}, denoted by $\mathcal{L}(\mathcal{H})$, which is an orthomodular lattice, i.e., a non-distributive generalization of Boolean algebra. Despite the lack of distributivity, the said algebraic structure is identified with a bivalent (though non-classical) semantics. In view of that, one might expect that algebraic structures identified with non-bivalent semantics should allow of rather drastic simplification of the KS proof.\\

\noindent The purpose of the present paper is to show that such a guess is correct. Particularly, the paper demonstrates that within the algebraic structure identified with supervaluationism – the form of a partial, non-bivalent semantics – in which the meet of two different nontrivial subspaces belonging to the different Boolean sub-algebras is undecidable, the impossibility of assigning truth values to all the experimental propositions, while preserving the truth-functional relations between them, becomes evident.\\

\section{Encoding the notion of superposition in logico-algebraic objects}  

\noindent Consider the set\smallskip

\begin{equation}  
   F 
   =
   \left\{
      x
      ,
      y
      \mkern3mu
      |
      \mkern3mu
      x
      \in
      y
   \right\}
   \;\;\;\;   ,
\end{equation}
\smallskip

\noindent where $x$ and $y$ are placeholders such that $x$ can be replaced by any pure quantum state of the system, and $y$ can be replaced by any closed linear subspace of the Hilbert space $\mathcal{H}$ associated with the system. If some pure state $|\Psi\rangle$ and closed linear subspace $\mathcal{P}$ are elements of the set $F$, then the statement $|\Psi\rangle \in \mathcal{P}$ is true. Hence, the generic statement $x \in y$ can be referred to as \emph{the predicate} or the propositional function $\mathfrak{P}_{\in}(x,y)$ which may take on the value true or false depending on its arguments $x$ and $y$. Otherwise stated, $\mathfrak{P}_{\in}(x,y)$ can be presented as the image of a couple $(x,y)$ under the Boolean-valued function $\mathfrak{P}_{\in}$ denoted by\smallskip

\begin{equation}  
   \mathfrak{P}_{\in}
   :
   \mathbb{X}
   \times
   \mathbb{Y}
   \to
   \mathbb{B}_{2}
   \;\;\;\;   ,
\end{equation}
\smallskip

\noindent where $\mathbb{X}$ is the set of all pure quantum states of the system and $\mathbb{Y}$ is the set of all closed linear subspaces of the Hilbert space $\mathcal{H}$.\\

\noindent Given that each experimental quantum proposition, say $P$, is accordant with the closed linear subspace, say $\mathcal{P}$, one can say that the truth value of $P$ in the state $|\Psi\rangle$ is defined by $\mathfrak{P}_{\in}(|\Psi\rangle,\mathcal{P})$, namely,\smallskip

\begin{equation} \label{FORM} 
   \mathfrak{P}_{\in}
   \!
   \left(
      |\Psi\rangle
      ,
      \mathcal{P}
   \right)
   =
   {[\mkern-3.3mu[
      P
   ]\mkern-3.3mu]}_v
   \;\;\;\;  ,
\end{equation}
\smallskip

\noindent if the propositional function $\mathfrak{P}_{\in}$ is determined on the couple $(|\Psi\rangle,\mathcal{P})$, that is, the statement $|\Psi\rangle \in \mathcal{P}$ is either true or false.\\

\noindent Let $\mathcal{A}$ and $\mathcal{B}$ be closed linear subspaces of $\mathcal{H}$ having no element in common except the zero-subspace $\{0\}$ such that they decompose $\mathcal{H}$ into their direct sum:\smallskip

\begin{equation}  
   \mathcal{A} 
   \cap
   \mathcal{B}
   =
   \mathcal{A} 
   \wedge
   \mathcal{B}
   = 
   \{0\}
   \;\;\;\;  ,
\end{equation}
\\[-40pt]

\begin{equation}  
   \mathcal{A} 
   \oplus
   \mathcal{B}
   =
   \mathcal{A} 
   \vee
   \mathcal{B}
   = 
   \mathcal{H}
   \;\;\;\;  ,
\end{equation}
\smallskip

\noindent where $\cap$ denotes the set-theoretic intersection, while $\wedge$ and $\vee$ denote the lattice-theoretic meet and join, respectively. The subspaces $\mathcal{A}$ and $\mathcal{B}$ represent the experimental propositions $A$ and $B$ which have the truth value of 1 in their relating states $|\Psi_{A}\rangle$ and $|\Psi_{B}\rangle$, namely,\smallskip

\begin{equation}  
   \mathfrak{P}_{\in}
   \!
   \left(
      |\Psi_{A}\rangle
      ,
      \mathcal{A}
   \right)
   =
   {[\mkern-3.3mu[
      A
   ]\mkern-3.3mu]}_v
   =
   1
   \;\;\;\;  ,
\end{equation}
\\[-40pt]

\begin{equation}  
   \mathfrak{P}_{\in}
   \!
   \left(
      |\Psi_{B}\rangle
      ,
      \mathcal{B}
   \right)
   =
   {[\mkern-3.3mu[
      B
   ]\mkern-3.3mu]}_v
   =
   1
   \;\;\;\;  .
\end{equation}
\smallskip

\noindent Taking into consideration that any vector $|\Psi\rangle$ describing the state of the system belongs to $\mathcal{H}$ but does not reside in $\{0\}$, one finds\smallskip

\begin{equation}  
   \mathfrak{P}_{\in}
   \!
   \left(
      |\Psi\rangle
      ,
      \mathcal{A}
      \vee
      \mathcal{B}
   \right)
   =
   1
   \;\;\;\;  ,
\end{equation}
\\[-40pt]

\begin{equation}  
   \mathfrak{P}_{\in}
   \!
   \left(
      |\Psi\rangle
      ,
      \mathcal{A}
      \wedge
      \mathcal{B}
   \right)
   =
   0
   \;\;\;\;  .
\end{equation}
\smallskip

\noindent Providing $\mathcal{A}\vee\mathcal{B}$ and $\mathcal{A}\wedge\mathcal{B}$ represent the disjunction ${A}\sqcup{B}$ and the conjunction ${A}\sqcap{B}$, respectively, and assuming that ${[\mkern-3.3mu[\top]\mkern-3.3mu]}_v=1$ and ${[\mkern-3.3mu[\perp]\mkern-3.3mu]}_v=0$, where $\top$ and $\perp$ denote correspondingly arbitrary tautology and contradiction, one gets\smallskip

\begin{equation}  
   {[\mkern-3.3mu[
      A
      \sqcup
      B
   ]\mkern-3.3mu]}_v
   =
   {[\mkern-3.3mu[
      \top
   ]\mkern-3.3mu]}_v
   \;\;\;\;  ,
\end{equation}
\\[-40pt]

\begin{equation}  
   {[\mkern-3.3mu[
      A
      \sqcap
      B
   ]\mkern-3.3mu]}_v
   =
   {[\mkern-3.3mu[
      \perp
   ]\mkern-3.3mu]}_v
   \;\;\;\;  ,
\end{equation}
\smallskip

\noindent which implies that the propositions $A$ and $B$ are \emph{mutually exclusive}, i.e., ${A}\sqcap{B}=\perp$, and they cannot be false together, i.e., ${A}\sqcup{B}=\top$.\\

\noindent As the direct sum of the subspaces $\mathcal{A}$ and $\mathcal{B}$ is the set that is formed by taking linear combinations of vectors in these subspaces, one can write down:\smallskip

\begin{equation}  
   \mathcal{A} 
   \oplus
   \mathcal{B}
   =
   \Big\{
      |\Psi_{A}\rangle\in\mathcal{A}
      ,
      |\Psi_{B}\rangle\in\mathcal{B}
      ,
      \mkern5mu
      c_{A},c_{B}\in\mathbb{C}
      \textnormal{:}      
      \mkern10mu            
      c_{A}|\Psi_{A}\rangle
      +
      c_{B}|\Psi_{B}\rangle
   \Big\}
   \;\;\;\;  .
\end{equation}
\smallskip

\noindent Let the state $|\Psi_{C}\rangle$ be \emph{a superposition of the states $|\Psi_{A}\rangle$ and $|\Psi_{B}\rangle$}, namely, $|\Psi_{C}\rangle = c_{A}|\Psi_{A}\rangle + c_{B}|\Psi_{B}\rangle$, such that\smallskip

\begin{equation}  
   \mathfrak{P}_{\in}
   \!
   \left(
      |\Psi_{C}\rangle
      ,
      \mathcal{C}
   \right)
   =
   {[\mkern-3.3mu[
      C
   ]\mkern-3.3mu]}_v
   =
   1
   \;\;\;\;  ,
\end{equation}
\smallskip

\noindent where $C$ is the proposition represented by the closed linear subspace $\mathcal{C}$. Then, the statement $|\Psi_{C}\rangle \in \mathcal{A}\vee \mathcal{B}$ is true, i.e.,\smallskip

\begin{equation}  
   \mathfrak{P}_{\in}
   \!
   \left(
      |\Psi_{C}\rangle
      ,
      \mathcal{A}
      \vee
      \mathcal{B}
   \right)
   =
   1
   \;\;\;\;  ,
\end{equation}
\smallskip

\noindent even though the statements $|\Psi_{C}\rangle \in \mathcal{A}$ and $|\Psi_{C}\rangle \in \mathcal{B}$ are not true.\\

\noindent Such a behavior of quantum superposition can be interpreted as the failure of the distributive law in the logic of experimental quantum propositions, that is,\smallskip

\begin{equation} \label{DIS} 
   C 
   \sqcap
      \left(
         A
         \sqcup
         B
      \right)
   \neq 
   \left(
      C
      \sqcap
      A
   \right)
   \sqcup
   \left(
      C
      \sqcap
      B
   \right)
   \;\;\;\;  .
\end{equation}
\smallskip

\noindent Indeed, the left-hand side of (\ref{DIS}) is $C\sqcap\top = C$ which has the value of true in the state $|\Psi_{C}\rangle$. Had the distributive law held true, it would have entailed the truth of of either $A$ or $B$ in $|\Psi_{C}\rangle$, that is, a contradiction to the behavior of quantum superposition.\\

\noindent Therefore, in order to construct a logico-algebraic account of quantum mechanics, one may propose \emph{to modify the rules of classical reasoning}. Specifically, one may demand that the algebraic structure defined on the set of the closed linear subspaces of the Hilbert space $\mathcal{H}$ relaxes the distributive properties of conjunction and disjunction. This kind of an algebraic structure is a Hilbert lattice $\mathcal{L}(\mathcal{H})$ \cite{Pavicic, Redei}. The aforesaid proposal, whose development has started about 80 years ago (and is not finished yet), is called quantum logic \cite{Birkhoff, Mackey}.\\

\noindent Be that as it may, a superposition of states $|\Psi_{A}\rangle$ and $|\Psi_{B}\rangle$ may well be encoded into a completely different semantics. Indeed, from the fact that $c_{A}|\Psi_{A}\rangle + c_{B}|\Psi_{B}\rangle$ belongs to neither subspace $\mathcal{A}$ nor subspace $\mathcal{B}$, which decompose $\mathcal{H}$ into the direct sum $\mathcal{A}\oplus\mathcal{B} = \mathcal{H}$, one can infer that the logic of experimental quantum propositions has a semantics that admits non-bivalent truth values, i.e., ones that are neither true nor false (while it retains the meaning of classical logical connectives $\sqcup$ and $\sqcap$).\\

\noindent Such is either \emph{a ``gappy'' semantics} allowing truth-value gaps or \emph{a many-valued semantics} allowing more than two truth values. In both semantics, the valuational formula (\ref{FORM}) stays valid even when the statement $|\Psi\rangle \in \mathcal{P}$ is undetermined, i.e., neither true nor false.\\

\noindent To be sure, in \emph{supervaluationism} \cite{Varzi, Keefe}, i.e., a semantics admitting truth-value gaps, if $\mathfrak{P}_{\in}$ is undetermined on the couple $(|\Psi\rangle, \mathcal{P})$, then the proposition $P$ represented by $\mathcal{P}$ has no truth value at all. In this way, one gets\smallskip

\begin{equation}  
   {[\mkern-3.3mu[
      P
   ]\mkern-3.3mu]}_v
   =
   \left\{
      \begingroup\SmallColSep
      \begin{array}{r l}
         x \in \mathbb{B}_{2}
         ,
         &
         \mkern8mu
         |\Psi\rangle
         \in
         \mathcal{P}
         \mkern5mu
         \text{is either true or false}
         \\
         \\[-7pt]
         0/0
         ,
         &
         \mkern8mu
         |\Psi\rangle
         \in
         \mathcal{P}
         \mkern5mu
         \text{is neither true nor false}
      \end{array}
      \endgroup   
   \right.
   \;\;\;\;  ,
\end{equation}
\smallskip

\noindent where $0/0$ symbolizes an indeterminate value. Since $c_{A}|\Psi_{A}\rangle + c_{B}|\Psi_{B}\rangle$ belongs to neither summand of the direct sum $\mathcal{A}\oplus\mathcal{B} = \mathcal{H}$, the statement $(c_{A}|\Psi_{A}\rangle + c_{B}|\Psi_{B}\rangle) \in \mathcal{A}$ is neither true nor false; consequently, one finds $\mathfrak{P}_{\in}(\mkern2mu c_{A}|\Psi_{A}\rangle + c_{B}|\Psi_{B}\rangle,\mkern3mu \mathcal{A}) = {[\mkern-3.3mu[A]\mkern-3.3mu]}_v = 0/0$; the same holds for $B$.\\

\noindent Unlike supervaluationism, a many-valued semantics fills in truth-value gaps with different degrees of gapness or truth degrees. E.g., in the infinite-valued {\L}ukasiewicz logic \cite{Pykacz}, one has the {\L}ukasiewicz disjunction and conjunction of compatible propositions $P$ and $Q$:\smallskip

\begin{equation}  
   {[\mkern-3.3mu[
      P
      \sqcap
      Q
   ]\mkern-3.3mu]}_v
   =
   \max
   \left(
      {[\mkern-3.3mu[
         P
      ]\mkern-3.3mu]}_v
      +
      {[\mkern-3.3mu[
         Q
      ]\mkern-3.3mu]}_v
      -
      1
      \mkern3mu
      ,
      0
   \right)
   \;\;\;\;  ,
\end{equation}
\\[-35pt]

\begin{equation}  
   {[\mkern-3.3mu[
      P
      \sqcup
      Q
   ]\mkern-3.3mu]}_v
   =
   \min
   \left(
      {[\mkern-3.3mu[
         P
      ]\mkern-3.3mu]}_v
      +
      {[\mkern-3.3mu[
         Q
      ]\mkern-3.3mu]}_v
      \mkern3mu
      ,
      1
   \right)
   \;\;\;\;  ,
\end{equation}
\smallskip

\noindent therefore, the mutually exclusiveness of the propositions, e.g., ${P}\sqcap{Q}=\perp$, results in\smallskip

\begin{equation}  
   {[\mkern-3.3mu[
      P
   ]\mkern-3.3mu]}_v
   +
   {[\mkern-3.3mu[
      Q
   ]\mkern-3.3mu]}_v
   \le
   1
   \;\;\;\;  ,
\end{equation}
\smallskip

\noindent and so\smallskip

\begin{equation}  
   {[\mkern-3.3mu[
      P
   ]\mkern-3.3mu]}_v
   =
   \left\{
      \begingroup\SmallColSep
      \begin{array}{r l}
         x \in \mathbb{B}_{2}
         ,
         &
         \mkern8mu
         |\Psi\rangle
         \in
         \mathcal{P}
         \mkern5mu
         \text{is either true or false}
         \\
         \\[-7pt]
         x \in (0,1)
         ,
         &
         \mkern8mu
         |\Psi\rangle
         \in
         \mathcal{P}
         \mkern5mu
         \text{is neither true nor false}
      \end{array}
      \endgroup   
   \right.
   \;\;\;\;  .
\end{equation}
\smallskip

\noindent This implies $\mathfrak{P}_{\in}(\mkern2mu c_{A}|\Psi_{A}\rangle + c_{B}|\Psi_{B}\rangle,\mkern3mu \mathcal{A}) = {[\mkern-3.3mu[A]\mkern-3.3mu]}_v \in (0,1)$; the same holds for B.\\

\noindent It must be noted that for any ``gappy'' semantics, one may find a many-valued semantics which would define the same logic \cite{Beziau} and, thus, the same structure of closed linear subspaces of $\mathcal{H}$. However, a many-valued semantics is much more complex than a ``gappy'' semantics. Therefore, for the sake of simplicity and succinctness of the explanation, one may consider only an algebraic structure identified with a ``gappy'' semantics.\\

\section{Relation between algebraic structures and semantics}  

\noindent Let us analyze the relation between algebraic structures, which can be defined on mathematical representatives of experimental quantum propositions, and semantics of those propositions involving truth values.\\

\noindent Recall that a closed linear subspace of the Hilbert space $\mathcal{H}$, say $\mathcal{P}$, is the range of the corresponding projection operator, say $\hat{P}$, acting on $\mathcal{H}$ \cite{Kalmbach}, explicitly,\smallskip

\begin{equation}  
   \mathcal{P}
   =
   \mathrm{ran}(\hat{P})
   =
   \left\{
      |\Psi\rangle
      \in
      \mathcal{H}
      \textnormal{:}
      \mkern10mu
      \hat{P}
      |\Psi\rangle
      =
      |\Psi\rangle      
   \right\}
   \;\;\;\;  .
\end{equation}
\smallskip

\noindent Because the set of the eigenvalues of each $\hat{P}$ is contained in $\{0,1\}$, one can assume correspondence between an experimental proposition $P$ and a projection operator $\hat{P}$, which is another way of stating that the mathematical representative of an experimental proposition $P$ is a closed linear subspace $\mathrm{ran}(\hat{P})$.\\

\noindent As the dimensionality of the Hilbert space $\mathcal{H}$ for a given quantum system depends on the number of all mutually exclusive experimental propositions $A$, $B$, $C$, … relating to the system, one can define a set $\Sigma$ of such propositions, i.e.,\smallskip

\begin{equation}  
   \Sigma
   =
   \left\{
      A
      ,
      B
      ,      
      C
      ,
      \dots
   \right\}
   \;\;\;\;  ,
\end{equation}
\smallskip

\noindent  as \emph{a context}. Equally, the context can be defined as a set $\hat{\Sigma}$ of nontrivial (i.e., differ from the identity operator $\hat{1}$ and the zero operator $\hat{0}$) projection operators\smallskip

\begin{equation}  
   \hat{\Sigma}
   =
   \left\{
      \hat{A}
      ,
      \hat{B}
      ,      
      \hat{C}
      ,
      \dots
   \right\}
   \;\;\;\;   
\end{equation}
\smallskip

\noindent such that any two members of $\hat{\Sigma}$, say $\hat{A}$ and $\hat{B}$, are orthogonal to each other, i.e.,\smallskip

\begin{equation}  
   \hat{A}
   \hat{B}
   =
   \hat{B}
   \hat{A}
   =
   \hat{0}
   \;\;\;\;  ,
\end{equation}
\smallskip

\noindent and the resolution of identity is associated with $\hat{\Sigma}$:\smallskip

\begin{equation}  
   \hat{A}
   +
   \hat{B}
   +
   \hat{C}
   +
   \dots
   =
   \hat{1}
   \;\;\;\;  .
\end{equation}
\smallskip

\noindent A subspace $\mathcal{P} \subseteq \mathcal{H}$ is called \emph{invariant} under the projection operator $\hat{P}$ on $\mathcal{H}$ if the image of every vector $|\Psi\rangle$ in $\mathcal{P}$ under $\hat{P}$ remains within $\mathcal{P}$. In symbols, this can be written as $\hat{P}\mathcal{P}\subseteq\mathcal{P}$ or, explicitly,\smallskip

\begin{equation}  
   \hat{P}
   \mathcal{P}
   =
   \left\{
      |\Psi\rangle
      \in
      \mathcal{P}
      \textnormal{:}
      \mkern10mu
      \hat{P}
      |\Psi\rangle
      \in
      \mathcal{P}      
   \right\}
   \;\;\;\;  .
\end{equation}
\smallskip

\noindent For example, since $\hat{P} \{0\} \subseteq \{0\}$ and $\hat{P} \mathcal{H} \subseteq \mathcal{H}$, the subspaces $\{0\}$ and $\mathcal{H}$ are invariant under every $\hat{P}$.\\

\noindent Accordingly, one can introduce the set $\mathcal{L}(\hat{\Sigma})$ of the invariant subspaces that are \emph{invariant under every projection operator of the context $\hat{\Sigma}$}:\smallskip

\begin{equation}  
   \mathcal{L}(\hat{\Sigma})
   =
   \bigcap_{\hat{P} \in \hat{\Sigma}}
   \left\{
      \mathcal{P}
      \subseteq
      \mathcal{H}
      \textnormal{:}
      \mkern10mu
      \hat{P}
      \mathcal{P}
      \subseteq
      \mathcal{P}
   \right\}
   \;\;\;\;  .
\end{equation}
\smallskip

\noindent Elements of this set form a complete lattice called \emph{the invariant-subspace lattice of the context $\hat{\Sigma}$} \cite{Radjavi}. It is straightforward to verify that each invariant-subspace lattice $\mathcal{L}(\hat{\Sigma})$ contains only closed linear subspaces corresponding to mutually commuting projection operators, implying that each $\mathcal{L}(\hat{\Sigma})$ is a Boolean algebra.\\

\noindent The collection of the lattices $\mathcal{L}(\hat{\Sigma})$, which is in one-to-one correspondence with with the set $\hat{\mathcal{O}}$ of all the contexts $\hat{\Sigma}$ associated with the quantum system, can be defined as\smallskip

\begin{equation}  
   C(\hat{\mathcal{O}})
   =
   \Big\{
      \hat{\Sigma}
      \in
      \hat{\mathcal{O}}
      \textnormal{:}
      \mkern10mu
      \mathcal{L}(\hat{\Sigma})
   \Big\}
   \;\;\;\;  .
\end{equation}
\smallskip

\noindent To make things easier, imagine the set $\hat{\mathcal{O}}$ which does not comprise \emph{interlinked contexts}, i.e., contexts that involve projection operators belonging to two or more contexts. In that case, the lattices $\mathcal{L}(\hat{\Sigma})$ have no identical elements other than the subspaces $\{0\}$ and $\mathcal{H}$.\\

\noindent Suppose that nontrivial -- i.e., other than $\{0\}$ and $\mathcal{H}$ -- subspaces $\mathcal{P}$ and $\mathcal{Q}$, which represent elementary (atomic) propositions about the quantum system, belong to different lattices, say $\mathcal{L}(\hat{\Sigma}^{\mkern2mu\prime})$ and $\mathcal{L}(\hat{\Sigma}^{\mkern2mu\prime\prime})$ in that order. Now, consider the set-theoretic intersection of the subspaces $\mathcal{P}$ and $\mathcal{Q}$. Using the set builder notation, this intersection can be presented as follows:\smallskip

\begin{equation}  
   \mathcal{P}
   \cap
   \mathcal{Q}
   =
   \Big\{
      |\Psi\rangle
      \in
      \mathcal{H}
      \textnormal{:}
      \mkern10mu
      \mathfrak{P}_{\in}
      \!
      \left(
         |\Psi\rangle
         ,
         \mathcal{P}
      \right)
      \sqcap
      \mathfrak{P}_{\in}
      \!
      \left(
         |\Psi\rangle
         ,
         \mathcal{Q}
      \right)
      \mkern-3mu
   \Big\}
   \;\;\;\;  .
\end{equation}
\smallskip

\noindent Consequently, if the vector $|\Phi\rangle$ in $\mathcal{H}$ is an element of $\mathcal{P}\cap\mathcal{Q}$, then both $|\Phi\rangle\in\mathcal{P}$ and $|\Phi\rangle\in\mathcal{Q}$ must be true.\\

\noindent Let the system be in the state described by the vector $|\Psi_{\mathcal{P}}\rangle$ such that the statement $|\Psi_{\mathcal{P}}\rangle\in\mathcal{P}$ is determined, i.e., either true or false. Because $\mathcal{P}$ and $\mathcal{Q}$ belong to the different lattices, the statement $|\Psi_{\mathcal{P}}\rangle\in\mathcal{Q}$ cannot be true.\\

\noindent To show this, let us present the vector $|\Psi_{\mathcal{P}}\rangle$ as a superposition of the vectors $|\Psi_{\mathcal{Q}}\rangle$ and $|\Psi_{\mathcal{Q}^{\perp}}\rangle$, namely, $|\Psi_{\mathcal{P}}\rangle = c_1 |\Psi_{\mathcal{Q}}\rangle + c_2 |\Psi_{\mathcal{Q}^{\perp}}\rangle$, where the statements $|\Psi_{\mathcal{Q}}\rangle\in\mathcal{Q}$ and $|\Psi_{\mathcal{Q}^{\perp}}\rangle\in\mathcal{Q}^{\perp}$ are determined, $c_1$ and $c_2$ are complex coefficients, while $(\cdot)^{\perp}$ denotes the set of vectors orthogonal to all vectors in $(\cdot)$. Evidently, if $|\Psi_{\mathcal{P}}\rangle\in\mathcal{P}$ is determined, then neither $|\Psi_{\mathcal{P}}\rangle\in\mathcal{Q}$ nor $|\Psi_{\mathcal{P}}\rangle\in\mathcal{Q}^{\perp}$ can be true.\\

\noindent In a bivalent semantics, this implies the falsity of the statements $|\Psi_{\mathcal{P}}\rangle\in\mathcal{Q}$ and $|\Psi_{\mathcal{P}}\rangle\in\mathcal{Q}^{\perp}$ resulting in the conclusion that the vector $|\Psi_{\mathcal{P}}\rangle$ is not an element of $\mathcal{P}\cap\mathcal{Q}$. Likewise, if the state of the system is described by the vector $|\Psi_{\mathcal{Q}}\rangle$ that makes the statement $|\Psi_{\mathcal{Q}}\rangle\in\mathcal{Q}$ determined, $|\Psi_{\mathcal{Q}}\rangle$ is not an element of $\mathcal{P}\cap\mathcal{Q}$.\\

\noindent It follows then that any vector $|\Psi\rangle$, which describes the state of the system (that is, any non-zero vector $|\Psi\rangle$ in $\mathcal{H}$), cannot belong to $\mathcal{P}\cap\mathcal{Q}$, i.e., \smallskip

\begin{equation}  
   \mathfrak{P}_{\in}
   \!
   \left(
      |\Psi\rangle
      ,
      \mathcal{P}
      \cap
      \mathcal{Q}
   \right)
   =
   0
   \;\;\;\;  .
\end{equation}
\smallskip

\noindent This means that the statement $|\Psi\rangle\in\mathcal{P}\cap\mathcal{Q}$ corresponds to an arbitrary contradiction $\bot$, which indicates that\smallskip

\begin{equation}  
   \mathcal{P}
   \cap
   \mathcal{Q}
   =
   \{0\}
   \;\;\;\;  .
\end{equation}
\smallskip

\noindent Hence, the subspaces $\mathcal{P}$ and $\mathcal{Q}$ have \emph{common} lower bound, the zero-subspace $\{0\}$. Consequently, the lattices $\mathcal{L}(\hat{\Sigma}^{\mkern2mu\prime})$ and $\mathcal{L}(\hat{\Sigma}^{\mkern2mu\prime\prime})$ can be joined together at the subspace $\{0\}$.\\

\noindent In the same way, one finds that $\mathcal{P}^{\perp}\cap\mathcal{Q}^{\perp}=\{0\}$. Since $\mathcal{H}$ is the smallest closed subspace of $\mathcal{H}$ containing $\mathcal{P}\cup\mathcal{Q}=\{0\}^{\perp}=\mathcal{H}$, the subspaces $\mathcal{P}$ and $\mathcal{Q}$ have common upper bound, $\mathcal{H}$. Therefore, the lattices $\mathcal{L}(\hat{\Sigma}^{\mkern2mu\prime})$ and $\mathcal{L}(\hat{\Sigma}^{\mkern2mu\prime\prime})$ can be joined together at the subspace $\mathcal{H}$ as well.\\

\noindent In view of that, one can join (or paste) together all the lattices of the set $\hat{\mathcal{O}}$ at the subspaces $\{0\}$ and $\mathcal{H}$. In accordance with the conjecture of M. Dichtl \cite{Dichtl} (proved in \cite{Navara}), by doing so one gets the Hilbert lattice $\mathcal{L}(\mathcal{H})$. In symbols, this can be presented as the union of the collection $C(\hat{\mathcal{O}})$, i.e., the set of all the subspaces in the collection $C(\hat{\mathcal{O}})$:\smallskip

\begin{equation}  
   \mathcal{L}(\mathcal{H})
   =
   \bigcup
   C(\hat{\mathcal{O}})
   =
   \bigcup_{\hat{\Sigma} \in \hat{\mathcal{O}}}
   \mathcal{L}(\hat{\Sigma})
   \;\;\;\;  .
\end{equation}
\smallskip

\noindent Now, turn to an algebraic structure identified with a “gappy” semantics, specially, supervaluationism.\\

\noindent In accord with the supervaluational notions of \emph{super-truth} and \emph{super-falsity}, a proposition $P$ is considered super-true if the predicate $\mathfrak{P}_{\in}$ definitely applies to a relation among the vector $|\Psi\rangle$ and the subspace $\mathcal{P}$ (where $\mathcal{P}$ represents $P$). Together with that $P$ is regarded as super-false if the said predicate definitely does not apply to a relation among $|\Psi\rangle$ and $\mathcal{P}$.\\

\noindent The predicate $\mathfrak{P}_{\in}$ definitely applies to a relation among $|\Psi\rangle$ and $\mathcal{H}$ (in other words, the statement $|\Psi\rangle\in\mathcal{H}$ is true in any state of the system); however, $\mathfrak{P}_{\in}$ definitely does not apply to a relation among $|\Psi\rangle$ and $\{0\}$ (i.e., the statement $|\Psi\rangle\in\{0\}$ is false without regard to system’s state). So, super-truth and super-falsity can be equated with the elements $\mathcal{H}$ and $\{0\}$ of each lattice $\mathcal{L}(\hat{\Sigma})$.\\

\noindent In the semantics of supervaluationism, the fact that neither $|\Psi_{\mathcal{P}}\rangle\in\mathcal{Q}$ nor $|\Psi_{\mathcal{P}}\rangle\in\mathcal{Q}^{\perp}$ can be true, implies that the predicate $\mathfrak{P}_{\in}$ is neither applies nor does not apply to a relation among the vector $|\Psi_{\mathcal{P}}\rangle$ and the subspace $\mathcal{Q}$. Such a case (which can be called \emph{a borderline case}) constitutes the predicate's penumbra: in that case $\mathfrak{P}_{\in}$ has no value on the couple $(|\Psi_{\mathcal{P}}\rangle, \mathcal{Q})$; in symbols, $\mathfrak{P}_{\in}(|\Psi_{\mathcal{P}}\rangle, \mathcal{Q}) = 0/0$. The couple $(|\Psi_{\mathcal{Q}}\rangle, \mathcal{P})$ amounts to the predicate's penumbra too, namely, $\mathfrak{P}_{\in}(|\Psi_{\mathcal{Q}}\rangle, \mathcal{P}) = 0/0$.\\

\noindent Because of that, the values of the expressions such as $\mathfrak{P}_{\in}(|\Psi_{\mathcal{P}}\rangle, \mathcal{P}) \mkern4mu\sqcap\mkern4mu \mathfrak{P}_{\in}(|\Psi_{\mathcal{P}}\rangle, \mathcal{Q})$ and $\mathfrak{P}_{\in}(|\Psi_{\mathcal{Q}}\rangle, \mathcal{P}) \mkern4mu\sqcap\mkern4mu \mathfrak{P}_{\in}(|\Psi_{\mathcal{Q}}\rangle, \mathcal{Q})$ cannot be determined, which causes $|\Psi_{\mathcal{P}}\rangle\in\mathcal{P}\cap\mathcal{Q}$ and $|\Psi_{\mathcal{Q}}\rangle\in\mathcal{P}\cap\mathcal{Q}$ to be undetermined. Hence, for any non-zero vector $|\Psi\rangle$ in $\mathcal{H}$ one gets\smallskip

\begin{equation}  
   \mathfrak{P}_{\in}
   \!
   \left(
      |\Psi\rangle
      ,
      \mathcal{P}
      \cap
      \mathcal{Q}
   \right)
   =
   0/0
   \;\;\;\;  .
\end{equation}
\smallskip

\noindent This indicates that it is impossible to tell what element(s) of the lattices $\mathcal{L}(\hat{\Sigma}^{\mkern2mu\prime})$ and $\mathcal{L}(\hat{\Sigma}^{\mkern2mu\prime\prime})$ is (are) that the subspaces $\mathcal{P}$ and $\mathcal{Q}$ have in common. In other words, one cannot decide what element of those lattices is equal to $\mathcal{P}\cap\mathcal{Q}$.\\

\noindent In this way, the algebraic structure identified with the supervaluation semantics is a collection of the lattices $\mathcal{L}(\hat{\Sigma})$ (Boolean sub-algebras or blocks) in which the meet of two nontrivial (and nonidentical) subspaces belonging to the different blocks is \emph{undecidable}. For the sake of brevity, let us refer to this structure as $\mathcal{L}_{0/0}(\mathcal{H})$.\\

\section{The KS theorem in bivalent and non-bivalent semantics}  

\noindent The simplest examples of non-interlinked contexts $\hat{\Sigma}$ can be found in a two-dimensional Hilbert space $\mathbb{C}^{2}$ characterizing \emph{a qubit}, i.e., a two-state quantum system.\\

\noindent The atomic propositions about the qubit are ``The spin of the qubit along a given axis $Q\in\mathbb{R}^{3}$ is $\pm\frac{\hbar}{2}\mkern4mu$''; accordingly, these propositions can be replaced with the letters $Q_{\pm}$. Each of $Q_{\pm}$ is represented by $\mathrm{ran}(\hat{Q}_{\pm})$, the range of the projection operator $\hat{Q}_{\pm}$ to measure spin along the $Q$ axis either \emph{up} (denoted by $+$) or \emph{down} (denoted by $-$). The operators $\hat{Q}_{\pm}$ form the contexts $\hat{\Sigma}_{Q}$ of the qubit, namely,\smallskip

\begin{equation}  
   \hat{\Sigma}_{Q}
   =
   \left\{
      \hat{Q}_{+}
      ,
      \hat{Q}_{-}
   \right\}
   \;\;\;\;  .
\end{equation}
\smallskip

\noindent These contexts correspond to the invariant-subspace lattices (or Boolean blocks):\smallskip

\begin{equation}  
   \mathcal{L}
   \mkern-3mu
   \left(
      \mkern-3mu
      \hat{\Sigma}_{Q}
      \mkern-3mu
   \right)
   =
   \left\{
      \{0\}
      \mkern3mu
      ,
      \mkern3mu
      \mathrm{ran}(\hat{Q}_{+})
      \mkern3mu
      ,
      \mkern3mu
      \mathrm{ran}(\hat{Q}_{-})
      \mkern3mu
      ,
      \mkern3mu
      \mathbb{C}^{2}
   \right\}
   \;\;\;\;  .
\end{equation}
\smallskip

\noindent Consider the set $S=\{P_1,P_2\}$ whose elements are the statements about the qubit: ``The spin of the qubit along the $X$ axis is $+\frac{\hbar}{2}$ AND the spin of the qubit along the $Z$ axis is $\pm\frac{\hbar}{2}\mkern4mu$''; in symbols,\smallskip

\begin{equation}  
   P_1
   =
   X_{+}
   \sqcap
   Z_{+}
   \;\;\;\;  ,
\end{equation}
\\[-40pt]

\begin{equation}  
   P_2
   =
   X_{+}
   \sqcap
   Z_{-}
   \;\;\;\;  .
\end{equation}
\smallskip

\noindent In accordance with Birkhoff and von Neumann’ proposal \cite{Birkhoff}, the propositions $P_1$ and $P_2$ are represented by the closed linear subspaces of $\mathbb{C}^{2}$, namely, the set-theoretical intersections $\mathrm{ran}(\hat{X}_{+})\mkern1mu\cap\mkern2mu\mathrm{ran}(\hat{Z}_{+})$ and $\mathrm{ran}(\hat{X}_{+})\mkern1mu\cap\mkern2mu\mathrm{ran}(\hat{Z}_{-})$. Consistent with the definition of $P_1$ and $P_2$, their logical conjunction ${P_1}\sqcap{P_2}$ is represented by the intersection of three closed linear subspaces that can be unambiguously written as $\mathrm{ran}(\hat{X}_{+})\mkern1mu\cap\mkern2mu\mathrm{ran}(\hat{Z}_{+})\mkern1mu\cap\mkern2mu\mathrm{ran}(\hat{Z}_{-})$.\\

\noindent It is straightforward to demonstrate that in the Hilbert lattice $\mathcal{L}(\mathbb{C}^{2})$, i.e., the continuum of pastings of the sub-algebras $\mathcal{L}(\hat{\Sigma}_{Q})$\smallskip

\begin{equation}  
   \mathcal{L}(\mathbb{C}^{2})
   =
   \bigcup_{Q\in\mathbb{R}^{3}}
   \mathcal{L}
   \mkern-3mu
   \left(
      \mkern-3mu
      \hat{\Sigma}_{Q}
      \mkern-3mu
   \right)
   =
   \bigcup_{Q\in\mathbb{R}^{3}}
   \left\{
      \{0\}
      \mkern3mu
      ,
      \mkern3mu
      \mathrm{ran}(\hat{Q}_{+})
      \mkern3mu
      ,
      \mkern3mu
      \mathrm{ran}(\hat{Q}_{-})
      \mkern3mu
      ,
      \mkern3mu
      \mathbb{C}^{2}
   \right\}
   \;\;\;\;  ,
\end{equation}
\smallskip

\noindent the compatible propositions $P_1$ and $P_2$ have truth values satisfying the product and sum rules. Certainly, the only element of $\mathcal{L}(\mathbb{C}^{2})$ that any pair of the subspaces $\mathrm{ran}(\hat{X}_{+})$, $\mathrm{ran}(\hat{Z}_{+})$ and $\mathrm{ran}(\hat{Z}_{-})$ have in common is the zero-subspace $\{0\}$; hence, according to the valuational formula (\ref{FORM}), one finds that $P_1$ and $P_2$ are compatible and their truth values satisfy the product rule: $\mathfrak{P}_{\in}(|\Psi\rangle, \{0\}) = {[\mkern-3.3mu[{P_1}\sqcap{P_2}]\mkern-3.3mu]}_v = {[\mkern-3.3mu[{P_1}]\mkern-3.3mu]}_v = {[\mkern-3.3mu[{P_2}]\mkern-3.3mu]}_v = 0$. Besides, provided that in $\mathcal{L}(\mathbb{C}^{2})$ the disjunction ${P_1}\sqcup{P_2}$ is represented by\smallskip

\begin{equation}  
   \mathcal{P}_1
   \vee
   \mathcal{P}_2
   =
   \left(
      \mathcal{P}_1^{\perp}
      \cap
      \mathcal{P}_2^{\perp}
   \right)^{\perp}
   \mkern-6mu
   =
   \{0\}
   \;\;\;\;  ,
\end{equation}
\smallskip

\noindent where $\mathcal{P}_1$ and $\mathcal{P}_2$ refer to $\mathrm{ran}(\hat{X}_{+})\mkern1mu\cap\mkern2mu\mathrm{ran}(\hat{Z}_{+})$ and $\mathrm{ran}(\hat{X}_{+})\mkern1mu\cap\mkern2mu\mathrm{ran}(\hat{Z}_{-})$, respectively, one finds ${[\mkern-3.3mu[{P_1}\sqcup{P_2}]\mkern-3.3mu]}_v = 0$, which implies that $ {[\mkern-3.3mu[{P_1}]\mkern-3.3mu]}_v$ and ${[\mkern-3.3mu[{P_2}]\mkern-3.3mu]}_v$ also satisfy the sum rule.\\

\noindent Let’s examine the structure $\mathcal{L}_{0/0}(\mathbb{C}^2)$ of the closed linear subspaces of the Hilbert space $\mathbb{C}^2$. In this structure, the intersection $\mathrm{ran}(\hat{X}_{+})\mkern1mu\cap\mkern2mu\mathrm{ran}(\hat{Z}_{+})\mkern1mu\cap\mkern2mu\mathrm{ran}(\hat{Z}_{-})$ is the meet of the subspace $\mathrm{ran}(\hat{X}_{+})$, which belongs to the lattice $\mathcal{L}(\hat{\Sigma}_{X})$, and the zero-subspace $\mathrm{ran}(\hat{Z}_{+})\mkern1mu\cap\mkern2mu\mathrm{ran}(\hat{Z}_{-}) = \{0\}$ belonging to the lattice $\mathcal{L}(\hat{\Sigma}_{Z})$. However, the zero-subspace in the lattice $\mathcal{L}(\hat{\Sigma}_{X})$ is identical to the zero-subspace in the lattice $\mathcal{L}(\hat{\Sigma}_{Z})$. Hence, $\mathrm{ran}(\hat{X}_{+})\mkern1mu\cap\mkern2mu\mathrm{ran}(\hat{Z}_{+})\mkern1mu\cap\mkern2mu\mathrm{ran}(\hat{Z}_{-})$ is the meet of the subspaces belonging to the same block, which means that it is decidable. Specifically,\smallskip

\begin{equation}  
   \mathfrak{P}_{\in}
   \!
   \left(
      |\Psi\rangle
      ,
      \mathrm{ran}(\hat{X}_{+})
      \cap
      \{0\}
   \right)
   =
   {[\mkern-3.3mu[{P_1}\sqcap{P_2}]\mkern-3.3mu]}_v
   =
   0 
   \;\;\;\;  .
\end{equation}
\smallskip

\noindent From here it follows that within the structure $\mathcal{L}_{0/0}(\mathbb{C}^2)$, the propositions $P_1$ and $P_2$ remain compatible as they allow the attribution of the truth value to their logical conjunction $ {P_1}\sqcap{P_2}$.\\

\noindent With that, in this structure, the intersections $\mathrm{ran}(\hat{X}_{+})\mkern1mu\cap\mkern2mu\mathrm{ran}(\hat{Z}_{+})$ and $\mathrm{ran}(\hat{X}_{+})\mkern1mu\cap\mkern2mu\mathrm{ran}(\hat{Z}_{-})$ are undecidable because their subspaces are different, nontrivial and belong to the different lattices $\mathcal{L}(\hat{\Sigma}_{X})$ and $\mathcal{L}(\hat{\Sigma}_{Z})$. Thus,\smallskip

\begin{equation}  
   \mathfrak{P}_{\in}
   \!
   \left(
      |\Psi\rangle
      ,
      \mathrm{ran}(\hat{X}_{+})
      \mkern-1mu
      \cap
      \mkern2mu
      \mathrm{ran}(\hat{Z}_{+})
   \right)
   =
   \mathfrak{P}_{\in}
   \!
   \left(
      |\Psi\rangle
      ,
      \mathcal{P}_1
   \right)
   =
   {[\mkern-3.3mu[{P_1}]\mkern-3.3mu]}_v
   =
   0/0
   \;\;\;\;  ,
\end{equation}
\\[-40pt]

\begin{equation}  
   \mathfrak{P}_{\in}
   \!
   \left(
      |\Psi\rangle
      ,
      \mathrm{ran}(\hat{X}_{+})
      \mkern-1mu
      \cap
      \mkern2mu
      \mathrm{ran}(\hat{Z}_{-})
   \right)
   =
   \mathfrak{P}_{\in}
   \!
   \left(
      |\Psi\rangle
      ,
      \mathcal{P}_2
   \right)
   =
   {[\mkern-3.3mu[{P_2}]\mkern-3.3mu]}_v
   =
   0/0
   \;\;\;\;  .
\end{equation}
\smallskip

\noindent Allowing that any binary operation, which applies to two indeterminate values $0/0$, gives an indeterminate value $0/0$ again, one gets that the product rule ${[\mkern-3.3mu[{P_1}\sqcap{P_2}]\mkern-3.3mu]}_v = {[\mkern-3.3mu[{P_1}]\mkern-3.3mu]}_v \times {[\mkern-3.3mu[{P_2}]\mkern-3.3mu]}_v$ fails in the structure $\mathcal{L}_{0/0}(\mathbb{C}^2)$, namely, $0 \neq \mkern3mu 0/0$.\\

\noindent Now, consider the union $\mathcal{P}_1\cup\mathcal{P}_2$\smallskip

\begin{equation}  
   \mathcal{P}_1
   \cup
   \mathcal{P}_2
   =
   \Big\{
      |\Psi\rangle
      \in
      \mathcal{H}
      \textnormal{:}
      \mkern10mu
      \mathfrak{P}_{\in}
      \!
      \left(
         |\Psi\rangle
         ,
         \mathcal{P}_1
      \right)
      \sqcup
      \mathfrak{P}_{\in}
      \!
      \left(
         |\Psi\rangle
         ,
         \mathcal{P}_2
      \right)
      \mkern-3mu
   \Big\}
   \;\;\;\;  .
\end{equation}
\smallskip

\noindent As both $\mathfrak{P}_{\in}\!(\Psi\rangle,\mathcal{P}_1)$ and $\mathfrak{P}_{\in}\!(\Psi\rangle,\mathcal{P}_2)$ are undetermined, the expression $\mathfrak{P}_{\in}(|\Psi\rangle, \mathcal{P}_1) \mkern4mu\sqcup\mkern4mu \mathfrak{P}_{\in}(|\Psi\rangle, \mathcal{P}_2)$ cannot be determined either. Hence, $\mathcal{P}_1\cup\mathcal{P}_2$ is undecidable, and so the smallest closed subspace of $\mathbb{C}^2$ containing $\mathcal{P}_1\cup\mathcal{P}_2$ is undecidable too. Accordingly,\smallskip

\begin{equation}  
   \mathfrak{P}_{\in}
   \!
   \left(
      |\Psi\rangle
      ,
      \mathcal{P}_1
      \mkern-3mu
      \vee
      \mkern-2mu
      \mathcal{P}_2
   \right)
   =
   {[\mkern-3.3mu[{P_1}\sqcup{P_2}]\mkern-3.3mu]}_v
   =
   0/0
   \;\;\;\;  .
\end{equation}
\smallskip

\noindent It means that as long as the sum rule can be presented as ${[\mkern-3.3mu[{P_1}]\mkern-3.3mu]}_v + {[\mkern-3.3mu[{P_2}]\mkern-3.3mu]}_v - {[\mkern-3.3mu[{P_1}\sqcup{P_2}]\mkern-3.3mu]}_v = {[\mkern-3.3mu[{P_1}\sqcap{P_2}]\mkern-3.3mu]}_v$, one has the failure of this rule in the structure $\mathcal{L}_{0/0}(\mathbb{C}^2)$, namely, $0/0 \neq \mkern3mu 0$.\\

\noindent This finding coincides with the statement of the KS theorem, maintaining that there is a set $S$ of compatible experimental propositions associated with a quantum system such that it impossible to assign truth values to all propositions in $S$ in a manner, in which the truth values will satisfy the product and sum rules.\\

\section{Concluding remarks}  

\noindent If two states of the quantum system are $|\Psi_{A}\rangle\mkern-3mu\in\mkern-3mu\mathcal{A}$ and $|\Psi_{B}\rangle\mkern-3mu\in\mkern-3mu\mathcal{B}$, where $\mathcal{A}$ and $\mathcal{B}$, the closed linear subspaces of the Hilbert space $\mathcal{H}$, decompose $\mathcal{H}$ into their direct sum, then all other states of the system $|\Psi_{C}\rangle\mkern-2mu\in\mkern-2mu\mathcal{C}$ such that $\mathcal{C}\le\mathcal{A}\vee\mathcal{B}$ (where the ordering relation $\le$ corresponds to the set-inclusion $\subseteq$) are called superpositions of $|\Psi_{A}\rangle$ and $|\Psi_{B}\rangle$. Provided subspaces $\mathcal{A}$, $\mathcal{B}$ and $\mathcal{C}$ represent experimental propositions $A$, $B$ and $C$, while $\mathcal{A}\vee\mathcal{B}$ represents the disjunction ${A}\sqcup{B}$, one can observe that the truth of $C$ excludes that of $A$ and it excludes that of $B$.\\

\noindent Remarkably, from the standpoint of the propositional language, the notion of superposition can be encoded in various logico-algebraic objects. Therefore, the logic of experimental quantum propositions may have a range of possible semantics. That is, it might have either a total semantics, or a partial semantics (in which the valuation relation – i.e., a mapping from the set of atomic propositions to the set of two objects, 1 and 0 – is not total), or a many-valued semantics (in which the gap between 1 and 0 is completed with truth degrees). The problem is that it is not possible to put the logic of experimental quantum propositions to the test and see what semantics it really has. Empirically, one can prove logical conjunction only if it joins propositions which are represented by the closed linear subspaces corresponding to mutually commuting projection operators. As a result, one cannot determine by means of experiment or experience whether, for example, the conjunctions ${C}\sqcap{A}$ and ${C}\sqcap{B}$ are always false or whether they have no truth value at all.\\

\noindent Hence, one could, theoretically, organize closed linear subspaces of $\mathcal{H}$ in different structures – i.e., not only in a Hilbert lattice (or some its generalizations) identified with the bivalent semantics of quantum logic, but also in the algebraic structure identified with the supervaluation semantics where the meet of two nontrivial and nonidentical subspaces belonging to the different blocks is undecidable.\\

\noindent On the other hand, the latter structure can allow of rather drastic simplification of some no-go theorems in the foundation of quantum mechanics. As it has been demonstrated in the present paper, within this structure, the KS theorem, which asserts the impossibility of assigning pre-existent bivalent truth values to all experimental propositions about the quantum system, becomes quite evident.\\

\section*{Acknowledgement}  

\noindent The author owes the anonymous referee a huge debt of gratitude for the incisive yet constructive comments which made possible to extensively improve this paper.\\

\bibliographystyle{References}

\end{document}